\documentclass{article}
\usepackage{amstext}
\usepackage[dvips]{graphicx}

\begin{document}

\begin{titlepage}
\rightline{GATC-99-01}
\vspace{1in}
\centerline{\LARGE{\bf{Boundary TBA Equations for a}}} 
\centerline{\LARGE{\bf{Non-diagonal Theory}}}          
\vspace{0.5in}
\centerline{\LARGE{J.N.Prata}}
\vspace{0.15in}        
\centerline{\large{July, 1999}}       
\vspace{0.15in} 
\centerline{\large{Grupo de Astrof\'\i sica e Cosmologia}}
\centerline{\large{Departamento de F\'isica}}       
\centerline{\large{Universidade da Beira Interior}}
\centerline{\large{R. Marqu\^es d'Avila e Bolama }}
\centerline{\large{6200 Covilh\~a-Portugal}}
\vspace{0.15in}        

\centerline{\it \large{jprata@mercury.ubi.pt}}
\vspace{0.5in}
\begin{abstract}
We compute the boundary entropies for the allowed boundary conditions of the $SU(2)$-invariant principal chiral model at level $k=1$. We used the reflection factors determined in a previous work. As a by-product we obtain some miscellaneous results such as the ground-state energy for mixed boundary conditions as well as the degeneracies of the Kondo model in the underscreened and exactly screened cases. All these computations are in perfect agreement with known results.
\end{abstract}

\end{titlepage}

\section{Introduction}

Boundary integrable models in $(1+1)$-dimensions have attracted some
attention in recent years, especially in view of their successful
application to quantum impurity problems. The typical examples are the Kondo
model \cite{12}, dissipative quantum mechanics \cite{36}, quantum Hall
liquids with constriction \cite{37} and the Callan-Rubakov model \cite{21}.
The common minimal denominator in these situations is the fact that the bulk
theory is conformally invariant and it is the boundary that is responsible
for the broken scale invariance. Here, our purpose is to consider the
alternative situation, where the boundary respects the conformal invariance
of the theory and the renormalization group (RG) trajectory is controlled by
a bulk perturbation. The model in question is the $SU(2)$ principal chiral
model (PCM) at level $k=1$. In a previous work \cite{27}, we proposed the
set of permissible boundary conditions suggested by the symmetries of the
problem and computed the corresponding minimal reflection factors that are
compatible with the bulk scale invariant limit (Kondo problem). In this
work, we wish to go one step further and study the finite size effects both
in the infrared (IR) and ultraviolet (UV) limits, in analogy with the bulk
problem \cite{10}. This is done by boundary thermodynamic Bethe ansatz (TBA) 
\cite{17}-\cite{19}, \cite{26}. To motivate our conjecture, we compute the
boundary entropies for the \textit{quasi-particles} of the Kondo model in
the \textit{exactly screened} and \textit{underscreened} cases. This has
already been done previously \cite{43}-\cite{48}. However, this explicit
computation will be used as a means of comparison for further
generalizations.

For one of the boundary conditions the boundary degeneracy is shown to be
noninteger. Furthermore, the typical decrease of entropy along the RG flow
from the unstable UV fixed point to a more stable IR one is found to be
respected. This $g$-theorem was conjectured for theories that are scale
invariant in the bulk \cite{20}. However, it also seems to hold here.

Finally, we use our results to conjecture the form of the ground state
energy for mixed boundary conditions in the IR limit. It is found to be in
perfect agreement with boundary conformal field theory predictions \cite{24}.

\section{The principal chiral model and the Kondo effect}

The integrability of the PCM has been studied by various authors \cite{1}-%
\cite{11}, \cite{15}, \cite{16}. The RG analysis reveals that it
interpolates between two fixed points. The crossover between the two
limiting behaviours introduces a mass scale that breaks the conformal
invariance. The RG trajectory terminates at the IR fixed point where the
theory becomes massless at all distances. It is characterized by a conformal
field theory based on two $SU(2)_k$ Kac-Moody algebras (at level $k$).
Al.B.Zamolodchikov and A.B.Zamolodchikov \cite{10} proposed the background
factorized scattering in terms of massless particles that leads to the
correct TBA equations for $k=1$. Zamolodchikov\'{}s $c$-function was shown
to take the values $c_{UV}=3$ and $c_{IR}=1$ at the fixed points \cite{10}, 
\cite{35}. The latter coincides with the central charge of the $SU(2)_1$
conformal field theory \cite{9}.

The multi-channel Kondo model (\cite{12}, \cite{13}, \cite{38}-\cite{48}),
on the other hand, consists of a $k$-tuple of $(1+1)$ free massless fermions
on the half-line antiferromagnetically coupled to a fixed impurity of spin $%
S $ sitting at the boundary $(x=0)$\footnote{%
A summary of the results presented here can be found in ref.\cite{12}.}. Let
us denote the coupling constant by $\lambda $. The RG flow interpolates
between an unstable UV fixed point, where the impurity is decoupled $%
(\lambda =0)$ and a strongly coupled IR one, where the spin of the impurity
is \textit{screened} $(\lambda =\frac{2}{k+2})$. The effective spin thus
becomes $q=S-k/2$. As before, the crossover introduces a scale $T_{K}$
called the Kondo temperature, \cite{39}-\cite{42}.

For $k=1$, the bulk spectrum of both theories consists of stable massless
particles: left- and right-movers. At level $k=1$ the kink structure of the
particles and boundary impurity is eroded. The on-mass-shell momenta of the
particles are parametrised in terms of the rapidity variables $- \infty<
\beta, \beta^{\prime}< \infty$: 
\begin{equation}
\left\{ 
\begin{array}{ll}
E=p= \frac{M}{2} e^{\beta}, & \mbox{for right-movers}, \\ 
&  \\ 
E=-p= \frac{M}{2} e^{- \beta^{\prime}}, & \mbox{for left-movers}.
\end{array}
\right.
\end{equation}
The right- and left-movers are represented by the symbols $R_a (\beta)$ and $%
L_{\bar b} (\beta^{\prime})$, respectively, where $a, \bar b = \pm$ are the $%
SU(2)$ isotopic spin indices.

The $SU(2)$ invariant $R$-$R$ scattering \cite{10} is given by: 
\begin{equation}
R_a (\beta_1) R_b (\beta_2) = S_{ab}^{cd} ( \beta_1 - \beta_2) R_d (\beta_2)
R_c ( \beta_1),
\end{equation}
with 
\begin{equation}
S_{ab}^{cd} ( \beta) = \sigma_T (\beta) \delta_a^c \delta_b^d + \sigma_R
(\beta) \delta_a^d \delta_b^c.
\end{equation}
The transition $(\sigma_T (\beta))$ and reflection amplitudes $(\sigma_R
(\beta))$ satisfy: 
\begin{equation}
\left\{ 
\begin{array}{l}
\sigma_T (\beta) = \frac{i}{\pi} \beta \sigma_R (\beta), \\ 
\\ 
\sigma_R (\beta) = - \frac{i \pi}{\beta - i \pi} S_V (\beta),
\end{array}
\right.
\end{equation}
where 
\begin{equation}
S_V (\beta) \equiv \sigma_T (\beta) + \sigma_R (\beta) = \frac{\Gamma \left( 
\frac{1}{2} + \frac{\beta}{2 i \pi} \right)\Gamma \left( - \frac{\beta}{2 i
\pi} \right)}{\Gamma \left( \frac{1}{2} - \frac{\beta}{2 i \pi} \right)
\Gamma \left( \frac{\beta}{2 i \pi} \right)},
\end{equation}
is the 2-particle amplitude in the isovector channel. All the previous
formulae hold for the $L$-$L$ scattering as well.

In the Kondo model the $R$-$L$ scattering is trivial. However, for the PCM,
we have: 
\begin{equation}
R_a ( \beta) L_{\bar b} (\beta^{\prime}) = U (\beta - \beta^{\prime})
L_{\bar b} ( \beta^{\prime}) R_a (\beta),
\end{equation}
with 
\begin{equation}
U (\beta) = \tanh \left( \frac{\beta}{2} - \frac{i \pi}{4} \right).
\end{equation}
Notice that when $\beta \to \pm \infty$, this amplitude becomes trivial.

Let us now consider a reflecting boundary that preserves the integrability.
In the Kondo problem we distinguish two situations \cite{13}. In the exactly
screened case, the impurity spin is completely screened yielding a zero
effective spin $(q=0)$. The impurity is thus a $SU(2)$ singlet. In the
underscreened case $(S=1)$, the effective spin is $q=1/2$ and the impurity
becomes a $SU(2)$ doublet.

In the former case, we define the operator $B$ which, acting on the vacuum $%
|0>$, creates a boundary state $|B>$ \cite{23}, i.e.: 
\begin{equation}
|B>=B|0>.
\end{equation}
The reflection matrix $R_{a}^{b}(\beta )$ is the amplitude corresponding to
a right-mover of rapidity $\beta $ and spin $a$ being reflected into a
left-moving state of rapidity $-\beta $ and spin $\bar{b}$\footnote{%
Sum over repeated indices is understood.}: 
\begin{equation}
R_{a}(\beta )B=R_{a}^{b}(\beta )L_{\bar{b}}(-\beta )B.
\end{equation}
The fact that the boundary impurity is a $SU(2)$ singlet implies the
following diagonal form: 
\begin{equation}
R_{a}^{b}(\beta )=\delta _{a}^{b}R_{K}(\beta ),
\end{equation}
where 
\begin{equation}
R_{K}(\beta )=U(\beta )=\tanh \left( \frac{\beta }{2}-\frac{i\pi }{4}\right)
.
\end{equation}
Similarly, for the underscreened case, we introduce the operator $B_{b}$ $%
(b=\pm )$, which acting on the vacuum creates a boundary state with spin $b$%
. The reflection amplitude $R_{ab}^{cd}(\beta )$ is: 
\begin{equation}
R_{a}(\beta )B_{b}=R_{ab}^{cd}(\beta )L_{\bar{c}}(-\beta )B_{d},
\end{equation}
with 
\begin{equation}
R_{ab}^{cd}(\beta )=f_{K}(\beta )\delta _{a}^{c}\delta _{b}^{d}+g_{K}(\beta
)\delta _{a}^{d}\delta _{b}^{c},
\end{equation}
and 
\begin{equation}
\left\{ 
\begin{array}{l}
f_{K}(\beta )=\frac{i}{\pi }\beta g_{K}(\beta ), \\ 
\\ 
g_{K}(\beta )=\sigma _{R}(\beta ).
\end{array}
\right.
\end{equation}
Here it is understood that $\beta $ stands for $\beta _{Q}-\beta _{K}$,
where $\beta _{Q}$ is the rapidity of the quasi-particle and $\beta _{K}$ is
associated with the Kondo temperature, $T_{K}=e^{\beta _{K}}$.

In the case of the PCM, we know that in the IR limit there are two boundary
conditions compatible with the conformal symmetry \cite{24}, \cite{25}. Each
one of them is associated with a primary field of the $SU(2)_1$ conformal
field theory. These fields have conformal dimensions $\Delta=0$ and $%
\Delta=1/4$ and have been identified with the identity operator and the
field $g$ of the Wess-Zumino-Witten action, respectively \cite{14}.

The '\textit{fixed}' boundary condition is associated with a reflection
amplitude of the form (9), (10) except that this time \cite{27}: 
\begin{equation}
R_P (\beta) = \exp \left(- \frac{i \pi}{ 4} \right) \frac{\sinh \left( \frac{%
\beta}{2} - \frac{i \pi}{8} \right)}{\sinh \left( \frac{\beta}{2} + \frac{i
\pi}{8} \right)}.
\end{equation}
Similarly, there is an amplitude $R_{ab}^{cd} (\beta)$ associated with the '%
\textit{free}' boundary condition, which is given by (12), (13), (14) and
this time: 
\begin{equation}
g_P ( \beta) = i R_P (\beta) \sigma_R (\beta).
\end{equation}

\section{TBA in the \textit{L}-channel}


\subsection{Fixed boundary conditions}


We start by considering a system of \textit{N} right-movers with rapidities $%
\beta _{1},\cdots ,\beta _{N}$ and \textit{M} left-movers with rapidities $%
\beta _{1}^{\prime },\cdots ,\beta _{M}^{\prime }$ in an interval of length 
\textit{R} with \textit{fixed} boundary conditions on both ends. Since the
reflection matrix is diagonal on both sides, we can assume that the Bethe
wave function vanishes at the two extremes of the interval. We can thus
impose a ''\textit{standing wave quantization condition}'' \cite{18}.
Mathematically, this condition can be expressed in the form: 
\begin{equation}
\begin{array}{c}
exp(2iRp_{k})R^{2}(\beta _{k})\frac{1}{U(2\beta _{k})}\prod_{l=1}^{N}U(\beta
_{k}+\beta _{l})\prod_{j=1}^{M}U(\beta _{k}-\beta _{j}^{\prime })\times \\ 
\\ 
\times \sum_{a,b}T_{a}^{b}(\beta _{k}|\left\{ \beta \right\} _{1\cdots
N})\times \bar{T}_{\bar{b}}^{\bar{a}}(-\beta _{k}|\left\{ \beta ^{\prime
}\right\} _{1\cdots M})=1,\qquad k=1,\cdots ,N.
\end{array}
\end{equation}
In the above equations it is understood that $a=\bar{a}$ and $b=\bar{b}$. $%
R(\beta )$ is the reflection amplitude in eqs.(11) or (15). The term $\frac{1%
}{U(2\beta _{k})}$ arises because the particle does not interact with
itself. The transfer matrix $T_{a}^{b}(u|\left\{ \beta \right\} _{1\cdots
N}) $ is the $2^{N}\times 2^{N}$ matrix defined by: 
\begin{equation}
T_{a}^{b}(u)_{b_{1}\cdots b_{N}}^{c_{1}\dots
c_{N}}=S_{ab_{1}}^{a_{1}c_{1}}(u-\beta
_{1})S_{a_{1}b_{2}}^{a_{2}c_{2}}(u-\beta _{2})\cdots
S_{a_{N-1}b_{N}}^{bc_{N}}(u-\beta _{N})
\end{equation}
and is represented in fig.1.
\begin{figure}
\begin{center}
\includegraphics[width=8cm]{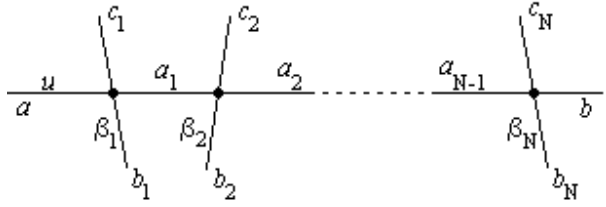}
\caption{Transfer Matrix}
\end{center}
\end{figure}

Each dot represents an interaction described by the amplitude (3).
Following the method described in ref.\cite{10}, we start by defining the
pseudo-energies $\varepsilon _{n}$ and the functions $L_{n}$ in the
thermodynamic limit, 
\begin{equation}
\begin{array}{l}
\frac{\rho _{0}}{\Lambda _{0}}=\frac{e^{-\epsilon _{0}}}{1+e^{-\epsilon _{0}}%
}, \\ 
\\ 
\frac{\tilde{\Lambda}_{n}}{\Lambda _{n}}=\frac{e^{-\epsilon _{n}}}{%
1+e^{-\epsilon _{n}}},\qquad n=1,2,\cdots ,\infty , \\ 
\\ 
L_{n}=\log \left( 1+e^{-\varepsilon _{n}\left( \beta \right) }\right) ,\
\;\;n=0,1,...,\infty ,\text{ }
\end{array}
\end{equation}

where $\rho _{n}$ is the density of $n$-magnons, $\Lambda _{n}$ the density
of $n$-magnon states and $\widetilde{\Lambda }_{n}=\Lambda _{n}-\rho _{n}$
is the density of holes. The scattering matrices for particle-magnon $\left(
S_{n}\right) $ and magnon-magnon $\left( S_{nm}\right) $ interactions are:

\begin{equation}
\begin{array}{l}
S_{n}(y-\beta )=\frac{y-\beta -in\pi /2}{y-\beta +in\pi /2}, \\ 
\\ 
S_{nm}(y)=\frac{y+i\pi (m+n)/2}{y-i\pi (m+n)/2}\times \left[ \frac{y+i\pi
(m+n-2)/2}{y-i\pi (m+n-2)/2}\cdots \frac{y+i\pi (m-n-2)/2}{y-i\pi (m-n-2)/2}%
\right] ^{2}\times \frac{y+i\pi (m-n)/2}{y-i\pi (m-n)/2},
\end{array}
\end{equation}
with kernels: 
\begin{equation}
\begin{array}{c}
\phi _{n}(y)=-i\frac{\partial }{\partial y}\log S_{n}(y), \\ 
\\ 
\phi _{nm}(y)=-i\frac{\partial }{\partial y}\log S_{nm}(y).
\end{array}
\end{equation}

The standard procedure leads to the set of Bethe equations:

\begin{equation}
\begin{array}{l}
-LMe^{\beta }+\epsilon _{0}+\frac{1}{2\pi }\varphi *L_{\bar{0}}+\frac{1}{%
2\pi }\varphi *L_{1}=0, \\ 
\\ 
\epsilon _{n}+\frac{1}{2\pi }\varphi *\sum_{m=0}^{\infty
}l_{mn}L_{m}=0,\qquad n=1,2,\cdots ,\infty ,
\end{array}
\end{equation}

where

\begin{equation}
L_{\bar{0}}(\beta )=L_{0}(-\beta ),
\end{equation}
and $l_{mn}$ $\left( m,n=0,1,...,\infty \right) $ is called the incidence
matrix. Its elements vanish except for adjacent nodes ( fig.2).
\begin{figure}
\begin{center}
\includegraphics[width=8cm]{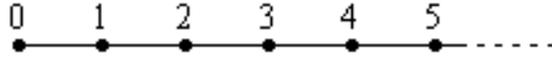}
\caption{Incidence Matrix}
\end{center}
\end{figure}
The unified kernel $\varphi \left( y\right) $ is given by:

\begin{equation}
\varphi (y)=\frac{1}{\cosh y}.
\end{equation}
The asymptotic values $x_{n}\equiv e^{-\varepsilon _{n}\left( -\infty
\right) }$, $y_{n}\equiv e^{-\varepsilon _{n}\left( +\infty \right) }$ are
given by:

\begin{eqnarray}
1+x_{n} &=&\frac{\sin ^{2}\left[ \frac{\pi (n+2)}{p+4}\right] }{\sin
^{2}\left[ \frac{\pi }{p+4}\right] },\qquad n=0,1,\cdots ,p, \\
1+y_{n} &=&\frac{\sin ^{2}\left[ \frac{\pi (n+1)}{p+3}\right] }{\sin
^{2}\left[ \frac{\pi }{p+3}\right] },\qquad n=0,1,\cdots ,p.
\end{eqnarray}
The infinite system has been truncated for some $p$. Eventually, we shall
take the limit $p\rightarrow \infty $.

The saddle point evaluation of the boundary free energy yields:

\begin{equation}
\log g^{\left( 1\right) }=\frac{1}{4\pi }\int_{-\infty }^{+\infty }d\beta
\Theta (\beta )L_{0}(\beta )+const,
\end{equation}

where

\begin{equation}
g^{(1)}=<0|B>
\end{equation}

is the boundary degeneracy and

\begin{equation}
\Theta (\beta )\equiv \frac{1}{i}\frac{\partial }{\partial \beta }\log
R^{2}(\beta )-\frac{1}{i}\frac{\partial }{\partial \beta }\log U(2\beta ).
\end{equation}
In the IR limit, when $L\rightarrow \infty $, we see from (22) that $%
\varepsilon _{0}\rightarrow \infty $ and so the first term on the right-hand
side of eq.(27) vanishes. In the UV limit $\left( L\rightarrow 0\right) $ we
get: 
\begin{eqnarray}
\log (g_{UV}^{(1)}) &=&\frac{1}{2}\log (1+e^{-\epsilon _{0}(-\infty
)})+const= \\
&=&\frac{1}{2}\lim_{p\rightarrow \infty }\log \left[ \frac{\sin ^{2}(\frac{%
2\pi }{p+4})}{\sin ^{2}(\frac{\pi }{p+4})}\right] +const=\log 2+const.
\end{eqnarray}
Which means that:

\begin{equation}
\frac{g_{UV}^{(1)}}{g_{IR}^{(1)}}=2\qquad \left( \mbox{PCM}\right) .
\end{equation}
For the Kondo problem, the second term in eq.(29) vanishes and $R(\beta )$is
given by eq.(11). And so: 
\begin{equation}
\frac{g_{UV}^{(1)}}{g_{IR}^{(1)}}=\lim_{p\rightarrow \infty }\left[ \frac{%
\sin ^{2}(\frac{2\pi }{p+4})}{\sin ^{2}(\frac{\pi }{p+4})}\right] ^{\frac{1}{%
2}}=2\qquad \left( \mbox{Kondo}\right) .
\end{equation}
This is in perfect agreement with refs.\cite{43}-\cite{48}.

Before we conclude this section, let us make one remark. Notice that, from
(27), the magnons do not contribute to the boundary entropy. This is because
they account for the isotopic degrees of freedom. Since for fixed boundary
condition, the impurity is a singlet, the magnons will not be included. For
free boundary condition, on the other hand, the impurity becomes a doublet.
Therefore its spin degrees of freedom contribute to the boundary degeneracy.
We thus expect the magnons to play a role here. Indeed, this is what
happens, as we shall see in the next section.

\subsection{Free boundary conditions}


Let us assume that our system is now subject to free boundary condition at
the two extremes. We can assume that the standing wave condition (17) still
holds if we have free boundary conditions at the two extremes of the
interval. The following simple argument shows that this is indeed the case.

If the boundary satisfies the fixed boundary condition, the Bethe wave
function vanishes there. Consequently, it is reflected off the boundary with
a phase shift of $\pi$. The other boundary produces another shift of $\pi$,
bringing the Bethe wave function back to its original form. The end result
is the standing wave.

For a free boundary the Bethe wave function satisfies a Neumann boundary
condition corresponding to zero flux of momentum accross the boundary.
Consequently, it remains unscathed upon reflection (no phase shift).

In summary, after two reflections, the Bethe wave function finds itself in
its original shape both for free as well as fixed boundary conditions,
provided the same condition holds at the two boundaries.

Proceeding with the same program that followed eq.(17), we obtain the
following set of Bethe equations: 
\begin{equation}
\begin{array}{rll}
-LMe^{\beta }+ & \epsilon _{0}+\frac{1}{2\pi }\varphi *L_{\bar{0}}+\frac{1}{%
2\pi }\varphi *L_{1} & =0, \\ 
&  &  \\ 
& \epsilon _{n}+\frac{1}{2\pi }\varphi *\sum_{m=0}^{\infty }l_{mn}L_{m} & 
=0,\qquad n=1,2,\cdots ,\infty .
\end{array}
\end{equation}
The evaluation of the boundary free energy leads to

\begin{equation}
\mathcal{F}|_{boundary}=\frac{1}{2\pi }\int_{-\infty }^{+\infty }d\beta
\left\{ \Theta _{0}^{(2)}(\beta )L_{0}(\beta )+\Theta _{1}^{(2)}(\beta
)L_{1}(\beta )\right\} +const,
\end{equation}
where 
\begin{equation}
\begin{array}{l}
\Theta _{0}^{(2)}(\beta )=\Theta ^{(2)}(\beta )-\phi ^{B}(\beta ), \\ 
\Theta _{1}^{(2)}(\beta )=\varphi ^{B}(\beta ).
\end{array}
\end{equation}
For the PCM:

\begin{eqnarray}
\Theta _{P}^{(2)}(\beta ) &=&\frac{1}{i}\frac{\partial }{\partial \beta }%
\log h_{P}(\beta )-\frac{1}{i}\frac{\partial }{\partial \beta }\log U(2\beta
), \\
\phi _{P}^{B}(\beta ) &=&-i\frac{\partial }{\partial \beta }\log h_{P}(\beta
), \\
\varphi _{P}^{B}(\beta ) &=&\frac{1}{\cosh \beta }, \\
h_{P}(\beta ) &=&f_{P}(\beta )+g_{P}(\beta ).
\end{eqnarray}

We then have:

\begin{equation}
\frac{g_{UV}^{(2)}}{g_{IR}^{(2)}}(PCM)=\frac{(1+x_{0})^{\frac{1}{4}%
}(1+x_{1})^{\frac{1}{2}}}{(1+y_{1})^{\frac{1}{2}}}=\frac{3\sqrt{2}}{2}.
\end{equation}
Let us check that our formulae yield the correct results for the Kondo
model. First of all notice that $h_{K}(\beta )=S_{V}(\beta )$, and so:

\begin{equation}
\Theta _{K}^{(2)}(\beta )=-i\frac{\partial }{\partial \beta }\log
S_{V}(\beta )=\phi _{K}^{B}(\beta ).
\end{equation}

Consequently, from (36), the first term in the integrand of eq.(35)
vanishes. We are thus left with:

\begin{equation}
\frac{g_{UV}^{(2)}}{g_{IR}^{(2)}}\left( Kondo\right) =\frac{\left(
1+x_{1}\right) ^{\frac{1}{2}}}{(1+y_{1})^{\frac{1}{2}}}=\frac{3}{2}.
\end{equation}
These results are in perfect agreement with ref.\cite{43}-\cite{48} for the
underscreened case $q=S-k/2=1/2$.

\section{TBA in the $R$-Channel}

The previous work allows us to conjecture the set of TBA equations in the $R$%
-channel. In essence they are not very different from those in the $L$%
-channel. They arise from imposing periodic boundary conditions on all
Cooper pairs of the form $R_{a}(\beta )L_{\bar{b}}(-\beta )$, \cite{17}. If $%
\rho _{0}(\beta )$ is the density of pairs with rapidity in a small vicinity
around $\beta $, then the TBA equations for these pairs and the $n$-magnons
are: 
\begin{equation}
\begin{array}{rll}
-\nu _{0}+ & \epsilon _{0}+\frac{1}{2\pi }\varphi *L_{\bar{0}}+\frac{1}{2\pi 
}\varphi *L_{1} & =0, \\ 
&  &  \\ 
-\nu _{1}+ & \epsilon _{1}+\frac{1}{2\pi }\varphi *L_{0}+\frac{1}{2\pi }%
\varphi *L_{2} & =0, \\ 
&  &  \\ 
& \epsilon _{n}+\frac{1}{2\pi }\varphi *\sum_{m=1}^{\infty }\tilde{l}%
_{mn}L_{m} & =0;\qquad n=2,3,\cdots ,\infty .
\end{array}
\end{equation}
The matrix $\widetilde{l}_{mn}$ $\left( m,n=1,2,...,\infty \right) $ is
obtained from $l_{mn}$ by ommiting the first node. In the $L$-channel the
particles and magnons contribute with $\Theta _{0}(\beta )$ and $\Theta
_{1}(\beta )$, respectively, to the boundary degeneracies. This contribution
will translate to $\nu _{0}$ and $\nu _{1}$ in eq.(44). If $\Theta
_{i}(\beta )$, associated with a particular boundary condition $\alpha $, is
of the form $\frac{1}{i}\frac{\partial }{\partial \beta }logR_{\alpha
}(\beta )-\frac{1}{i}\frac{\partial }{\partial \beta }logS(2\beta )$, then
its contribution to $\nu _{i}(\beta )$ reads $-logK_{\alpha }(\beta )$,
where $K_{\alpha }(\beta )=R_{\alpha }(i\pi /2-\beta )$.

Suppose we have fixed boundary condition on one side and free on the other.
The contribution of the free boundary to $\nu _{1}$ is $-logK_{K}(\beta )$
where $K_{K}(\beta )=R_{K}(i\pi /2-\beta )$ is given by the Kondo reflection
amplitude (11). Its contribution to $\nu _{0}$ is $-logK_{P}(\beta )$, where 
$K_{P}(\beta )=R_{P}(i\pi /2-\beta )$ is given by eq.(15). In principle its
contribution should be $-logh_{P}(\beta )$. However, the amplitude in the
logarithm must be the one associated with the kernel in $\Theta _{0}^{(2)}$,
which was shown to be $R_{P}$ at the end of the previous section
(cf.eqs.(36), (42)). Finally, the fixed boundary does not contribute to $\nu
_{1}$ and its contribution to $\nu _{0}$ amounts to $-log\bar{K}_{P}(\beta )$%
.

Altogether, we have: 
\begin{equation}
\left\{ 
\begin{array}{l}
\nu _{0}(\beta )=RMe^{\beta }-log\left[ \bar{K}_{P}(\beta )K_{P}(\beta
)\right] , \\ 
\nu _{1}(\beta )=-logK_{K}(\beta ).
\end{array}
\right.
\end{equation}
Let us define $\lambda _{0}(\beta )=\bar{K}_{P}(\beta )K_{P}(\beta )$, $%
\lambda _{1}(\beta )=K_{K}(\beta )$, $\lambda _{n}(\beta )=1$, $n=2,3,\cdots
,\infty $ and 
\begin{equation}
\left\{ 
\begin{array}{l}
\tilde{\epsilon}_{n}(\beta )=\epsilon _{n}(\beta )+log\lambda _{n}(\beta ),
\\ 
\\ 
\tilde{L}_{n}(\beta )=log(1+\lambda _{n}(\beta )e^{-\tilde{\epsilon}_{n}}).
\end{array}
\right.
\end{equation}
We can then rewrite eq.(44) in the form: 
\begin{equation}
\begin{array}{rl}
-MRe^{\beta }+\tilde{\epsilon}_{0}+\frac{1}{2\pi }\varphi *\tilde{L}_{\bar{0}%
}+\frac{1}{2\pi }\varphi *\tilde{L}_{1} & =0, \\ 
&  \\ 
-m_{n}Re^{\beta }+\tilde{\epsilon}_{n}+\frac{1}{2\pi }\varphi
*\sum_{m=0}^{\infty }l_{mn}\tilde{L}_{m} & =0;\qquad n=1,2,\cdots ,\infty .
\end{array}
\end{equation}
Now notice that $\lambda _{0}(\beta )=\bar{K}_{P}(\beta )K_{P}(\beta )=1$
and $\lim_{\beta \to \pm \infty }\lambda _{1}(\beta )=\lim_{\beta \to \pm
\infty }K_{K}(\beta )=\mp 1$. If we define $z_{n}\equiv e^{-\tilde{\epsilon}%
_{n}(+\infty )}$, we get the following system in the limit $R\to \infty $: 
\begin{equation}
\left\{ 
\begin{array}{l}
z_{0}=0,\qquad z_{1}=1, \\ 
\\ 
1+z_{n}=\frac{sin^{2}\left[ \frac{\pi (n-1)}{p+1}\right] }{sin^{2}\left[ 
\frac{\pi }{p+1}\right] };\qquad n=2,3,\cdots ,p.
\end{array}
\right.
\end{equation}
Similarly, if we define $x_{n}\equiv e^{\tilde{\epsilon}_{n}(-\infty )}$, in
the region $-RM<<\beta $, we obtain the system (25), as $R\to \infty $.

The standard eavluation of the free energy yields: 
\begin{equation}
\mathcal{F}=-\frac{L}{2\pi R}\sum_{n=0}^{\infty }\left\{ \mathcal{L}\left[ 
\frac{x_{n}}{1+x_{n}}\right] -\mathcal{L}\left[ \frac{z_{n}}{1+z_{n}}\right]
\right\} ,
\end{equation}
where $\mathcal{L}(x)$ is Rodger\'{}s dilogarithm \cite{32}: 
\begin{equation}
\mathcal{L}(x)=-\frac{1}{2}\int_{0}^{x}dt\left[ \frac{logt}{1-t}+\frac{%
log(1-t)}{t}\right] .
\end{equation}
Using the sum rules of ref.\cite{32} we get for the ground state energy: $%
E_{0}(R)=-\frac{\mathcal{F}(R)}{L}=\frac{5\pi }{24R}.$

On the other hand\footnote{%
The central charge for the PCM at level $k=1$ is $c=1$.}: 
\begin{equation}
\Delta _{min}=\frac{R}{\pi }E_{0}(R)+\frac{c}{24}=\frac{1}{4},
\end{equation}
which is one of the conformal dimensions of the $SU(2)_{1}$ conformal
theory. This is in fact the correct one, according to the predictions of
boundary conformal field theory, \cite{24}, \cite{25}.

\section{Conclusions}

Let us restate our results. We computed the boundary degeneracies
corresponding to the two allowed boundary conditions for the PCM at level $%
k=1$. This was accomplished by quantizing the system in a finite box. We
argued that for the free boundary condition, the Bethe wave function was
subject to a Neumann condition preventing the flow of momentum accross the
boundary. This condition allows the exchange of isospin between the
particles and the boundary impurity. Since the magnons are attached to these
degrees of freedom, they contribute to the boundary degeneracy. This
situation is in contrast with the fixed boundary condition, which does not
permit isospin permutation. The free boundary condition was subsequently
shown to yield a noninteger degeneracy. As a consistency check, we compared
our results in the bulk scale invariant limit $\left( U(\beta )\rightarrow
1\right) $ with those of the Kondo model.

We stress that in all cases the boundary entropy decreases under
renormalization from the more unstable UV fixed point to the more stable IR
one.

Finally, we used the previous results to conjecture the form of the TBA
equations in the $R$-channel and the ground-state energy, when we let our
system evolve between two distinct boundary states. The boundary conformal
field theory approach of Cardy \cite{24} predicts a partition function of
the form $Z_{free,fixed}=\chi _{\frac{1}{2}}^{(1)}$ in the IR limit. $\chi _{%
\frac{1}{2}}^{(1)}$ denotes the $SU(2)$ Kac-Moody character at level $k=1$
associated with the conformal tower labeled by the primary operator with
isospin $l=\frac{1}{2}$. This operator has conformal dimension \cite{9} $%
\Delta =\frac{l(l+1)}{k+2}=\frac{1}{4}$, in perfect agreement with our
result (52). Some of our results may be tested numerically using the
truncated systems (25), (26) and (48). We expect to carry out that program
in a future work.

\vspace{0.5in}

\centerline{\large{Acknowledgements}}

\bigskip

The author wishes to thank P.Bowcock, E.Corrigan, P.Dorey and A.Taormina for
useful discussions and comments.



\begin{thebibliography}{10}
\bibitem[1]{1}  E.Witten, Commun. Math. 92 (1984) 455.

\bibitem[2]{2}  S.P.Novikov, Usp. Mat. Nauk. 37 (1982) 3.

\bibitem[3]{3}  M.C.M.Abdalla, Phys. Lett. B152 (1985) 215.

\bibitem[4]{4}  A.M.Polyakov, Phys. Lett. B72 (1979) 247.

\bibitem[5]{5}  Y.Y.Goldschmidt, E.Witten, Phys. Lett. B91 (1980) 392.

\bibitem[6]{6}  A.M.Polyakov, P.B.Wiegman, Phys. Lett. B141 (1984) 223.

\bibitem[7]{7}  A.M.Polyakov, P.B.Wiegman, Phys. Lett. B131 (1983) 121.

\bibitem[8]{8}  V.G.Knizhnik, A.B.Zamolodchikov, Nucl. Phys. B247 (1984) 83.

\bibitem[9]{9}  D.Gepner, E.Witten, Nucl. Phys. B278 (1986) 493.

\bibitem[10]{10}  A.B.Zamolodchikov, Al.B.Zamolodchikov, Nucl. Phys. B379
(1992) 602.

\bibitem[11]{11}  B.Berg, M.Karowski, P.Weizand, V.Kurak, Nucl. Phys. B134
(1979) 125.

\bibitem[12]{12}  I.Affleck, Lectures given at the XXXVth Cracow School of
Theoretical Physics, Zakopane, Poland, June 1995, Acta Polon. B26 (1995)
1869.

\bibitem[13]{13}  P.Fendley, Phys. Rev. Lett. 71 (1993) 2485.

\bibitem[14]{14}  A.B.Zamolodchikov, V.A.Fateev, Sov. J. Nucl. Phys. 43
(1986) 657.

\bibitem[15]{15}  A.N.Kirillov, F.A.Smirnov, Phys. Lett. B198 (1987) 506.

\bibitem[16]{16}  P.Mejean, F.A.Smirnov, Int. J. Mod. Phys. A12 (1997) 3383.

\bibitem[17]{17}  A.LeClair, G.Mussardo, H.Saleur, S.Skorik, Nucl. Phys.
B453 (1995) 581.

\bibitem[18]{18}  P.Fendley, H.Saleur, Nucl. Phys. B428 (1994)681.

\bibitem[19]{19}  P.Dorey, A.Pocklington, R.Tateo, G.Watts, Nucl. Phys. B525
(1998) 641.

\bibitem[20]{20}  I.Affleck, A.W.W.Ludwig, Phys. Rev. Lett. 67 (1991) 161.

\bibitem[21]{21}  I.Affleck, J.Sagi, Nucl. Phys. B417 (1994) 374.

\bibitem[22]{22}  T.R.Klassen, E.Melzer, Nucl. Phys. B338 (1990) 485.

\bibitem[23]{23}  S.Ghoshal, A.B.Zamolodchikov, Int. J. Mod. Phys. A9 (1995)
118.

\bibitem[24]{24}  J.L.Cardy, Nucl. Phys. B324 (1989) 581.

\bibitem[25]{25}  E.Verlinde, Nucl. Phys. B300 (1988) 360.

\bibitem[26]{26}  F.Lesage, H.Saleur, P.Simonetti, Phys. Lett. B427 (1998)
85.

\bibitem[27]{27}  J.N.Prata, Phys. Lett. B438 (1998) 115.

\bibitem[28]{28}  C.N.Yang, C.P.Yang, Phys. Rev. 150 (1966) 321.

\bibitem[29]{29}  C.N.Yang, Phys. Rev. Lett. 19 (1967) 1312.

\bibitem[30]{30}  M.Takahashi, Prog. Theor. Phys. 46 (1971) 401.

\bibitem[31]{31}  Al.B.Zamolodchikov, Nucl. Phys. B342 (1990) 695.

\bibitem[32]{32}  Al.B.Zamolodchikov, Nucl. Phys. B358 (1991) 497.

\bibitem[33]{33}  Al.B.Zamolodchikov, Nucl. Phys. B358 (1991) 524.

\bibitem[34]{34}  Al.B.Zamolodchikov, Phys. Lett. B253 (1991) 391.

\bibitem[35]{35}  Al.B.Zamolodchikov, Nucl. Phys. B366 (1991) 122.

\bibitem[36]{36}  A.J.Leggett, S.Chakravarty, A.T.Dorsey, M.P.A.Fisher,
A.Garg, W.Zwerger, Rev. Mod. Phys. 59 (1987) 1.

\bibitem[37]{37}  K.Moon, H.Yi, C.L.Kane, S.M.Kane, S.M.Girvin,
M.P.A.Fisher, Phys. Rev. Lett. 71 (1983) 4381.

\bibitem[38]{38}  N.Andrei, Phys. Rev. Lett. 45 (1980) 379.

\bibitem[39]{39}  J.Kondo, Prog.Theor. Phys. 32 (1964) 37.

\bibitem[40]{40}  K.G.Wilson, Rev. Mod. Phys. 47 (1975) 773.

\bibitem[41]{41}  P. Nozi\`{e}res, J. Low Temp. Phys. 17 (1974) 31.

\bibitem[42]{42}  P. Nozi\`{e}res, A.Blandin, J. Phys. 41 (1980) 41.

\bibitem[43]{43}  N.Andrei, K.Furuya, J.Lowenstein, Rev. Mod. Phys. 55
(1983) 331.

\bibitem[44]{44}  P.B.Wiegmann, Sov. Phys. J.E.T.P. Lett. 31 (1980) 392.

\bibitem[45]{45}  N.Andrei, C.Destri, Phys. Rev. Lett. 52 (1984) 364.

\bibitem[46]{46}  A.M.Tsvelick, P.B.Wiegmann, Z. Phys. B54 (1985) 364.

\bibitem[47]{47}  A.M.Tsvelick, P.B.Wiegmann, J. Stat. Phys. 38 (1985) 38.

\bibitem[48]{48}  A.M.Tsvelick, J. Phys. C18 (1985) 159.
\end{thebibliography}
\end{document}